\documentclass[epj]{webofc}
\usepackage[varg]{txfonts}   
\woctitle{QCD@Work 2014 }
\begin{document}

\title{Hybrid exotic mesons in soft-wall AdS/QCD }

\author{Loredana Bellantuono\inst{1,2}\fnsep\thanks{\email{loredana.bellantuono@ba.infn.it}}
}

\institute{INFN-Sezione di Bari, via Orabona 4, 70126 Bari, Italy \and
           Dipartimento di Fisica, Universit\`a degli Studi di Bari Aldo Moro, via Orabona 4, 70126 Bari, Italy
          }

\abstract{Hybrid mesons with exotic quantum numbers
$J^{PC}=1^{-+}$ are examined in soft-wall AdS/QCD. The predicted
mass spectrum is compared to the measured values of the candidates
$\pi_1(1400)$, $\pi_1(1600)$ and $\pi_1(2015)$. Thermal effects
are analysed through the spectral function in the AdS-Black Hole
model, and the differences with the Hawking-Page description are
discussed.}

\maketitle

\section{Introduction}
\label{intro}Quantum Chromodynamics (QCD) describes strong
interactions among quarks as processes in which colored gluons are
exchanged. By virtue of their own color charge, gluons also
strongly interact with quarks and among themselves. For this
reason, a meson should be considered as a linear superposition of
color-singlet bound states $
\left|q\overline{q}'\right\rangle,\left|qq\overline{q}'\overline{q}'\right\rangle,
\left|qG\overline{q}'\right\rangle, \left|GG\right\rangle,\dots, $
comprising quarks $(q)$, antiquarks $\left(\overline{q}\right)$
and gluons $(G)$ as constituents \cite{Barnes}. These states can
be classified, respectively, as quark model "quarkonia",
"multiquarks",  "hybrids", "glueballs", and so forth. These
structures determine angular momentum, parity and
charge-conjugation quantum numbers $J^{PC}$ of the meson, and they
yield also exotic combinations not included in the quark model
$q\overline{q}'$ picture.

Hybrid configuration, composed by a quark-antiquark pair plus a
constituent gluon, accounts for either ordinary or exotic $J^{PC}$
quantum numbers. Therefore, experimental evidence of these states
can come from overpopulations of the ordinary $J^{PC}$ spectra
compared to quark model prediction, or from the detection of exotic
states. The first strategy seems unfruitful, because of the
densely populated spectrum of light mesons in the mass region
between $1$ and $2$ GeV, and the broad nature of the states
involved \cite{Ketzer}. On the other hand, states that can have
exotic quantum numbers are multiquarks and hybrids. They have been
searched in experiments aiming at the detection of their decay
products, but the analysis has revealed cumbersome. The
identification of exotic resonances would be a strong argument
supporting the existence of hybrid bound states. Several QCD
models identify the meson with  $J^{PC}=1^{-+}$ as the
lowest-lying exotic state, with mass predictions varying between
$1.5$ and $2.2$ GeV \cite{Ketzer}. Currently, in the light quark
sector there are three candidates for hybrid $1^{-+}$ states:
$\pi_{1}(1400)$, $\pi_{1}(1600)$ and $\pi_{1}(2015)$. Their
measured masses are $M(\pi_1(1400))= 1354\pm 25$ MeV,
$M(\pi_1(1600))=1662^{+8}_{-9}$ MeV and $M(\pi_1(2015))=2014 \pm
20 \pm 16,2001 \pm 30 \pm 92$ MeV \cite{Beringer}. Further
information on the detection of such states can be found in the
bibliography of \cite{Bellantuono}. The nature of $1^{-+}$
candidates is still a debated issue, and new experiments with
higher statistics and better acceptance are expected to improve
present understandings \cite{Ketzer}.

I examine the $J^{PC}=1^{-+}$ mesons in the holographic approach, a
framework inspired by the AdS/CFT correspondence. This conjecture
relates type IIB string theory in an $AdS_5 \times S^5$ space
($AdS_5$ is a 5-dimensional anti-de Sitter space and $S^5$ is a
5-dimensional sphere) with $\cal{N}$=4 Super-Yang Mills theory in
a 4-dimensional Minkowski space \cite{Maldacena}. Holographic
duality prescribes the identification between the partition
function of the conformal field theory and the one of the string
theory on $AdS_5$ \cite{Witten253}. The
correspondence determines a dictionary \cite{Witten253,Gubser}
according to which a local gauge-invariant $p$-form with conformal
dimension $\Delta$ is dual to a 5-dimensional field with mass
$m_5$ given by $m_5^2 R^2=(\Delta-p)(\Delta+p-4)$, $R$ being the
curvature radius of the anti-de Sitter space. A holographic
description of QCD at finite temperature can be obtained by
modifying the metric on the $AdS$ space; I use this method for
discussing the stability of the hybrid configurations against
thermal fluctuations in comparison with other quark and gluon
bound states.

\section{Mass spectrum at $T=0$}
\label{Hybrid-mesons-null-T} In Poincar\'e coordinates the line
element of the anti-de Sitter space can be written as
\begin{equation}\label{eq:metric} ds^2=\frac{R^2}{z^2}(dt^2-d\vec
x^{\,2}-dz^2) \qquad\qquad z>0 \,,
\end{equation}
where $(t,\vec x\,)$ are the ordinary 4-dimensional coordinates, and
$z$ is the fifth holographic coordinate. The boundary of the
$AdS_5$ space is a compactification of Minkowski space
\cite{Witten253} and corresponds to the condition $z=0$. This
framework aims at reproducing QCD properties through a projection
on the boundary of the gravity theory defined in the $AdS$ space
(bulk). However, the AdS/CFT correspondence has been conjectured
for a conformal gauge theory, while QCD is not conformal
invariant. A bottom-up strategy for breaking this symmetry
requires the introduction of an infrared scale in the holographic
space. I adopt the soft-wall model, in which a background dilaton
field $\phi(z)=c^{2}z^{2}$ involving a dimensionful scale
$c=\mathcal{O}\left(\Lambda_{QCD}\right)$ implements a smooth
cutoff of the holographic space \cite{Karch,Andreev}; in the
so-called hard-wall model, the conformal symmetry is broken by
restricting the coordinate $z$ to a maximum value \cite{Erlich}.
The hybrid $1^{-+}$ meson can be represented by the QCD local
operator $J_\mu^a=\bar q T^a G^a_{\mu\nu} \gamma^\nu q$, with
$G^a_{\mu\nu}$ the gluon field strength tensor and $T^a$ flavour
matrices normalized to Tr$[T^aT^b]=\delta^{ab}/2$. The dual field
is a 1-form, $H_\mu=H_\mu^a T^a$, with mass  $m_5^2 R^2=8$ and
dynamics described by a Proca-like action:
\begin{equation}\label{eq:action}
S=\frac{1}{k}\int d^5x \, \sqrt{g}\, e^{-c^2z^2} \,
\mbox{Tr}\left[-\frac{1}{4} F^{MN}F_{MN} +\frac{1}{2} m_5^2 H_M
H^M  \right]\,,
\end{equation}
where $F_{MN}=\partial_M H_N-\partial_N H_M$, $M$ and $N$
5-dimensional indices, $g$ the determinant of the metric
(\ref{eq:metric}) and $k$ a parameter making the action
dimensionless. The mass spectrum and normalised eigenfunctions of
the $1^{-+}$ hybrid states
\begin{eqnarray}\label{eq:massspectrum}
q^2_n&=&M_n^2=4 c^2(n+2)   \,\,  ,\\
H_n&=&\sqrt{\frac{2\, n!}{(n+3)!}} \, c^4\, z^{4}\, L_n^3(c^2z^2) \,\,,
\,
\end{eqnarray}
with $n=0,1,2,\dots$ and $L_n^m$ the generalized Laguerre
polynomial, can be obtained from the Euler-Lagrange equation for
the transverse component of the bulk field $H^\mu_\perp$
$\left(\partial_\mu H^\mu_\perp=0\right)$ \cite{Bellantuono}.

\begin{figure}
 \centering
 \includegraphics[width=7cm,clip]{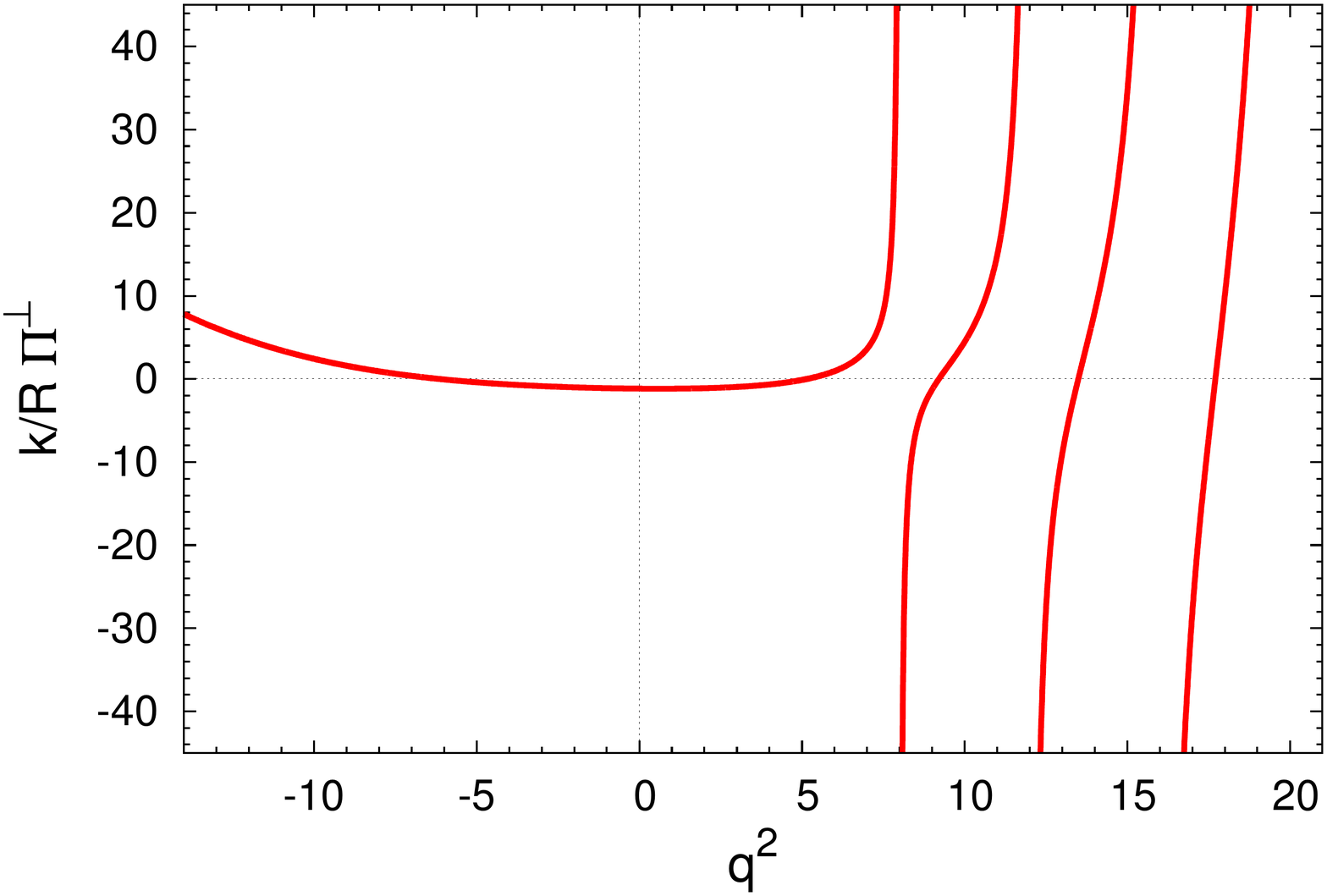}
 \caption{Regularized $\Pi^\perp(q^2)$ transverse invariant function; $c$ is set to one.}
 \label{fig:Piperp}
\end{figure}

Another analysis of the $1^{-+}$ meson properties can be performed
by means of the two-point correlation function of the local
operator $J_\mu^a$
\begin{equation}\label{eq:correlation}
\Pi^{ab}_{\mu\nu}= i \int d^4x \, e^{i q x} \langle 0|T[J^a_\mu(x)
J^b_\nu(0)]  |0\rangle \, =-\left(\eta_{\mu\nu}-\frac{q_\mu
q_\nu}{q^2}\right) \frac{\delta^{ab}}{2}\Pi^\perp(q^2)+\frac{q_\mu
q_\nu}{q^2} \frac{\delta^{ab}}{2} \Pi^\parallel(q^2)\,,
\end{equation}
where the decomposition in a transverse and a longitudinal
contribution has been used. Holographic QCD employs the weak
version of the AdS/CFT duality \cite{Edelstein}, relating the
strongly-coupled theory on the boundary to the supergravity theory
in the bulk. In this limit the identification between the dual
partition functions can be written as
\begin{equation}\label{eq:functional derivative}
\Pi_{\mu\nu}^{ab}(q^2)=\left. \frac{\delta^2 S_{os}}{\delta
H_0^{a\mu} \delta H_0^{b\nu}}\right|_{H_0=0}\,,
\end{equation}
where $S_{os}$ is the on-shell supergravity action
\eqref{eq:action} and $H_0^\mu(q^2)$ is the source, related to the
hybrid field $H^\mu(z,q)=H(z,q^2)H_0^\mu(q^2)$ by the
bulk-to-boundary propagator $H(z,q^2)$. A comparison between the
expressions \eqref{eq:correlation} and \eqref{eq:functional
derivative} allows one to determine the transverse function
$\Pi^\perp(q^2)$, whose poles coincide with the mass spectrum
\eqref{eq:massspectrum}, as shown in fig.~\ref{fig:Piperp}
\cite{Bellantuono}.
The residues can be calculated as well.

The soft-wall mass spectra of the various QCD mesonic bound states
comprising quarks and gluons (Table~\ref{tab-1}) are Regge
trajectories $M_n^2 \simeq n$, having all the same slope.
\begin{table}[b]
\centering \sidecaption \caption{Mass spectrum of $q \bar q$,
glueballs and hybrids predicted in the soft-wall model.}
\label{tab-1}
\begin{tabular}{lllll}
\hline $J^{PC}$ & $1^{--}$  $(q \bar q)$ \cite{Karch} \, &
$0^{++}$ $(q \bar q)$ \cite{Colangelo2008} & $0^{++}$ (glueball)
\cite{Colangelo2007} & $1^{-+}$ \cite{Bellantuono} \\\hline
$M^2_n$
& $c^2(4 n+4)$ & $c^2(4 n+6)$ & $c^2(4 n+8)$ & $c^2(4 n+8)$\\
\hline
\end{tabular}
\end{table}
The mass of the lowest-lying hybrid state is $M_0 \sim 1.1-1.3$
GeV, depending on the way the parameter $c$ is fixed from the
$1^{--}$ spectrum \cite{Karch, Andreev}. The  mass is lower than
the predictions obtained by other approaches \cite{Ketzer}, and
suggests $\pi_{1}(1400)$ as the lightest $1^{-+}$ meson. Moreover,
the mass spectrum \eqref{eq:massspectrum} gives an intriguing
description of the radial excited $1^{-+}$ states, as it can be
deduced  comparing the prediction $ M_1/M_0= \sqrt{3/2} \sim 1.22
$ with the experimental ratio $M(\pi_1(1600))/M(\pi_1(1400)) \sim
1.22$. According to this observation, one can interpret
$\pi_1(1400)$ and $\pi_1(1600)$ as the lowest-lying and the
first-excited state of the $1^{-+}$ spectrum, respectively. In the
hard-wall model, where the spectrum behaves as $M_n^2 \simeq n^2$,
the mass of the lowest-lying state is closer to $M(\pi_{1}(1400))$
than the soft-wall prediction, but the first radial excitation is
heavier \cite{Kim}.

\section{Stability against thermal effects}
\label{Hybrid-mesons-finite-T}The spectral function $\rho(q^2)$ of
$\Pi^\perp$ in (\ref{eq:correlation}), represented at zero
temperature by an infinite number of delta functions centered at
the eigenvalues of the mass spectrum, can
 be evaluated at increasing temperature in order to examine
in-medium effects on hadron properties.

\begin{figure}[b]
 \centering
 \includegraphics[width=7cm,clip]{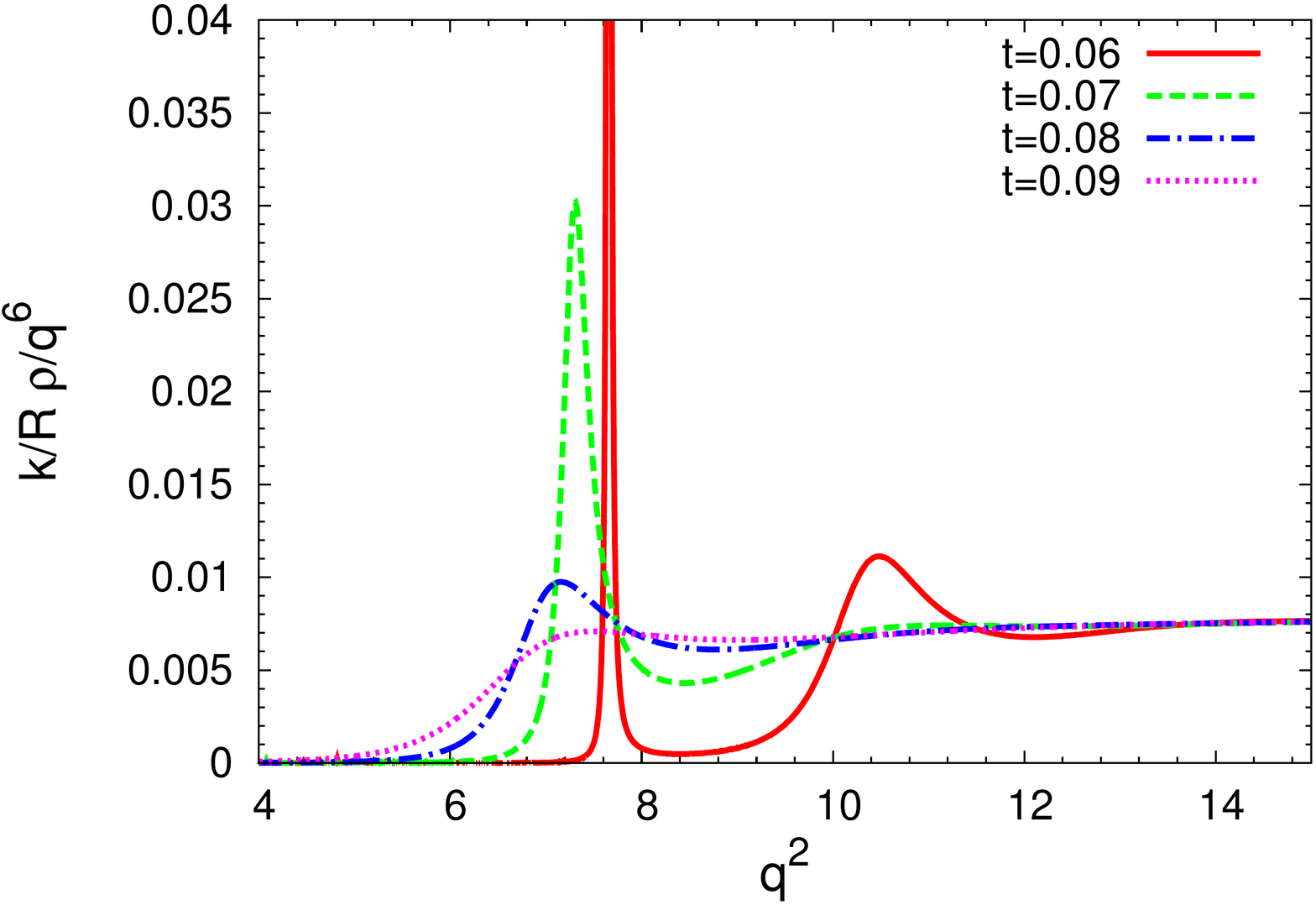}
 \caption{Spectral function of  $1^{-+}$ mesons at different values of temperature, with $c=1$.}
 \label{fig:spfunc}
\end{figure}

A possible framework for performing this analysis employs the
AdS-Black Hole metric
\begin{equation}\label{eq:metricBH}
ds^2=\frac{R^2}{z^2}\left[\left( 1-z^4/z_h^4 \right)dt^2-d\vec
x^{\,2}-\frac{dz^2}{1-z^4/z_h^4}\right] \qquad\qquad 0<z<z_h \,,
\end{equation}
with the position of its horizon $z_h$ depending on the
temperature, $z_h=1/(\pi T)$. From the action \eqref{eq:action}
and the geometry \eqref{eq:metricBH}, the bulk-to-boundary
propagator $H^{\perp}(z,q^2)$  can be computed in the meson
rest-frame, $\vec{q}=0$. The resulting spectral function is shown
in fig.~\ref{fig:spfunc} for a few values of the  temperature
$t=T/c$ \cite{Bellantuono}.
The broadening of the peaks and their shift towards lower values
of the squared mass when the temperature is increased, is visible.
A melting temperature, indicating the dissociation of the hybrid
states, can be identified from fitting the lowest-lying peak of
the spectral function with a Breit-Wigner formula
\cite{Bellantuono}. Table~\ref{tab-2} collects the computed
melting temperatures of various hadronic states.
\begin{table}[b]
\centering \sidecaption \caption{Melting temperature of  $q \bar
q$, glueball and hybrid states predicted in the soft-wall model.}
\label{tab-2}
\begin{tabular}{lllll}
\hline $J^{PC}$&$1^{--}$  $(q \bar q)$ \cite{Mamani} \,&$0^{++}$
$(q \bar q)$ \cite{Colangelo2009}&$0^{++}$ (glueball)
\cite{Colangelo2009}&$1^{-+}$ \cite{Bellantuono}  \\\hline
 $ t=T/c$& $0.23$&$0.18$&$0.12$&$0.10$ \\\hline
\end{tabular}
\end{table}
Thermal dissociation can also be described by means of a
quark-gluon binding potential, which gets a shape preventing the
formation of hybrid bound states at a critical temperature similar
to the one inferred from the spectral function \cite{Bellantuono}.
The analyses show that hybrid mesons melt at a lower temperature
compared to light vector and scalar mesons, and scalar glueballs.
In energy units, the hybrid $1^{-+}$ melting temperature
$T_{melting}$ is quite low, and  depends on the way $c$ has been
fixed,  ranging approximately from $40$ to $50$ MeV.

A deconfinement phase transition can be introduced in order to
describe the thermal response of the spatial string tension, which
is temperature independent below $\sim 210$ MeV and rapidly
increasing above this critical value \cite{Andreev2}. Another way
of performing the analysis of deconfinement includes the
Hawking-Page transition between a low-temperature confined phase
described by Thermal $AdS$ metric and a high-temperature
deconfined phase described by $AdS$ metric with black hole
\cite{Hawking,Witten505}. The transition occurs at the temperature
$T_{HP}$ ($\sim 190$ MeV in the soft-wall model) for which the
difference between the free energies of the two geometries changes
sign \cite{Herzog}. Since the confined-phase spectral function
equals the zero-temperature one, and $T_{HP}$ is considerably
higher than $T_{melting}$, the Hawking-Page scheme describes
thermal dissociation as a discontinuous process. Hadronic states
retain their zero-temperature spectrum despite thermal
fluctuations in the confined phase and suddenly disappear from the
spectrum (i.e. they melt) at deconfinement.

\section{Conclusions}
In the soft-wall  model  the hybrid spectrum seems to reproduce
the measured mass of the $1^{-+}$ candidates. Hybrids are found to
suffer of larger thermal instabilities with respect to other hadrons,
and this could explain the difficulties in their detection.
Further efforts are needed to develop a more refined dual model of
finite temperature QCD, in which the nontrivial AdS-Black Hole
picture of thermal dissociation holds at the more reasonable
Hawking-Page temperature scale.
\begin{acknowledgement}
I thank P. Colangelo and F. Giannuzzi for collaboration, and F.
Br\"{u}nner for fruitful discussions.
\end{acknowledgement}

\end{document}